# Gas-Liquid Nucleation in Two Dimensional System

**Mantu Santra, Suman Chakrabarty and Biman Bagchi**

Solid State and Structural Chemistry Unit,

Indian Institute of Science, Bangalore 560012, India.

*Abstract*

We study the nucleation of the liquid phase from a supersaturated vapor in two dimensions (2D). Using different Monte Carlo simulation methods, we calculate the free energy barrier for nucleation, the line tension and also investigate the size and shape of the critical nucleus. The study is carried out at an intermediate level of supersaturation (away from the spinodal limit). In 2D, a large cut-off in the truncation of the Lennard-Jones (LJ) potential is required to obtain converged results, whereas low cut-off (say, 2.5σ is generally sufficient in three dimensional studies, where σ is the LJ diameter) leads to a substantial error in the values of line tension, nucleation barrier and characteristics of the critical cluster. It is found that in 2D, the classical nucleation theory (CNT) fails to provide a reliable estimate of the free energy barrier. It *underestimates* the barrier by as much as 70% at the saturation-ratio S=1.1 (defined as S=P/$P_C$, where $P_C$ is the coexistence pressure at reduced temperature $T^*$= 0.427). Interestingly, CNT has been found to overestimate the nucleation free energy barrier in three dimensional (3D) systems near the triple point. In fact, the agreement with CNT is worse in 2D than in 3D. Moreover, the existing theoretical estimate of the line tension overestimates the value significantly.

## I. Introduction

Nucleation of the liquid phase from a supersaturated vapor phase is an activated process that involves the formation of a critical nucleus of the thermodynamically stable liquid-like embryo within the metastable vapor phase. The free energy of formation of the critical nucleus is determined by the free energy gain due to the formation of the liquid droplet and the free energy loss due to the creation of the gas-liquid interface. The classical nucleation theory (CNT)[1-4] uses these two terms to obtain an expression for the free energy of the formation of a liquid nucleus of radius $r$ which is given, in two dimensions, by

$$\Delta\Omega(r) = -\pi r^2 |\Delta\Omega_A| + 2\pi r \gamma_\infty \tag{1}$$

where $\Delta\Omega_A$ is the free energy difference per unit area between the stable liquid and the metastable gas phase, and $\gamma_\infty$ is the line tension of the planar interface. Here, the line tension for a nucleus of radius $r$ is assumed to be that of a planar surface between bulk phases using the capillary approximation. The maximum of $\Delta\Omega(r)$ gives the critical size of the nucleus ($r^*$) and the free energy barrier ($\Delta\Omega^*$) as given by the following expressions

$$r^* = \frac{\gamma_\infty}{\Delta\Omega_A}, \tag{2}$$

$$\Delta\Omega^* = \frac{\pi \gamma_\infty^2}{\Delta\Omega_A}. \tag{3}$$

The use of these expressions requires a priori knowledge about the free energy difference between the liquid and the vapor phase, and the line tension. While the former is available from the equation of state, it is somewhat harder to obtain a precise estimate of the line tension, although semi-quantitative estimates are often available.

The validity of Becker-Döring expression of the free energy barrier has been widely tested for gas-liquid nucleation in three dimensional (3D) systems, by carrying out both detailed theoretical and computer simulation studies. It is found that in 3D, the CNT free energy barrier overestimates the actual barrier by about 4-5 $k_BT$. However, such studies have hardly been carried out for 2D gas-liquid nucleation.

Two dimensional fluids are important not only because of theoretical interest, but they have practical relevance as well. Although perfect 2D fluids may not exist in nature, there are many systems that can be modeled well as a 2D system, like physisorption of gases on solid surface[5], surfactant monolayer adsorbed on an air/water interface[6], thinning of completely wetted films on clean mica surfaces[7] and many more. As already mentioned (and this point cannot be over emphasized) that despite the great importance of nucleation in 2D, there seem to be very few studies of the problem. An elegant microscopic theory developed by Zeng[8] and based on density functional approach of Zeng and Oxtoby[9], suggests that CNT should be more accurate for 2D than for 3D. This prediction has not yet been tested against simulations.

The organization of the rest of the paper is as follows. In the next section, we present a brief summary of the grand canonical transition matrix Monte Carlo (GC-TMMC) method used to compute the coexistence line and line tension. In section III we describe the model system and the simulation details. Section IV contains the results and their discussion. Section V contains concluding remarks.

## II. Grand Canonical Transition Matrix Monte Carlo (GC-TMMC) method

Grand canonical transition matrix Monte Carlo[11] is one of the methods used to get free energy by computer simulation of fluid phase. Simulations are conducted in grand canonical ensemble (volume V, chemical potential μ and temperature T are held constant) and TMMC method[10] is used to compute the free energy as a function of any reaction coordinate of choice. An outline of the TMMC method applied to the present problem follows.

The basic idea is to calculate the probability ($\Pi(S,\beta)$) that the system is in a macrostate S (in our case, size of a liquid cluster or density of the system) at the inverse temperature β (=1/$k_B$T, $k_B$ is Boltzmann constant and T is the temperature). Note that the average value of any observable of the system can be computed once the macrostate probability is known, e.g. in canonical ensemble, the free energy can be obtained as a function of order parameter from $\beta F(S,\beta) = -\ln \Pi(S,\beta)$. The *Boltzmann macrostate probability* can be expressed as $\Pi(S,\beta) = \sum_{s \in S} \pi(s,\beta)$, where $\pi(s,\beta)$ is the Boltzmann probability for a particular microstate (microscopic configuration) s. It is given by $\pi(s,\beta) = \exp(-\beta H_s)$ where $H_s$ is the value of the Hamiltonian for the microstate s.

The algorithm of finding $\Pi(S,\beta)$ is as follows:

1. Similar to Metropolis algorithm, for a given initial microstate 's', a new state 't' is proposed with probability $q_{s,t}$. As a simplification, we chose $q_{s,t} = q_{t,s}$ though it is not strictly necessary.

2. The probability of the move to state 't' being accepted is

$$r_{s,t}(\beta,\eta) = \min\left\{1, \frac{\exp(\eta_T)\pi(t,\beta)}{\exp(\eta_S)\pi(s,\beta)}\right\},$$

where $\eta_S$ is the weight function corresponding to macrostate S. For our purpose, we have set $\eta_S = -\ln\Pi(S,\beta)$, which corresponds to the *multicanonical approach*. Thus the macrostates with lower probability are given more weight so that towards the end of the simulation the histogram becomes flat and low probability states are sampled quite well.

3. A new bookkeeping step is incorporated following the equilibration of the above Markov chain. At every step an array $C_{S,T}$ (initialized to zero) is incremented as follows:

For $S \neq T$,

$$C_{S,T}(\beta) = C_{S,T}(\beta) + r_{s,t}(\beta,\eta=0)$$

$$C_{S,S}(\beta) = C_{S,S}(\beta) + (1 - r_{s,t}(\beta,\eta=0))$$

For $S = T, C_{S,S} = C_{S,S} + 1$. Note that while the Markov chain is guided by the multicanonical weight, the unweighted Boltzmann transition probabilities are stored for each visited macrostate.

The *canonical transition probability* (CTP) between the macrostates has been calculated at some interval as

$$P_{S,T}(\beta) = \frac{C_{S,T}(\beta)}{\sum_U C_{S,U}(\beta)}.$$

The equilibrated Markov chain must obey the detailed balance equation $\Pi_S(\beta)P_{S,T}(\beta) = \Pi_T P_{T,S}(\beta)$. Hence, the macrostate probabilities have been obtained by solving the set of coupled linear equations iteratively. The weights have been updated at every $1000 \times N$ steps ($N$ is the average number of particles in the system at metastable state in case of cluster size as the order parameter, whereas in case of density we use the number of particles at the maximum value of density) and the simulations have been continued for a long time until when the system samples the whole order parameter range with equal probability and the free energy converges to the true value. The system learned to pass through the low probability states automatically and reasonably good sampling was attained.

## III. Model system and simulation details

The model system studied consists of circular plate in which the particles interact with each other according to Lennard-Jones potential,

$$\Phi(r) = 4\varepsilon \left[ \left(\frac{\sigma}{r}\right)^{12} - \left(\frac{\sigma}{r}\right)^{6} \right]$$

where '$\varepsilon$' is the potential depth, '$\sigma$' is the diameter of a particle and '$r$' is the distance between two particles.

In our simulations we have studied systems both with above potential truncated and shifted at different radii ($r_c$ = 2.5$\sigma$, 4.0$\sigma$ and 7.0$\sigma$) and without any truncation. The truncated and shifted potential is given by,

$$\Phi_{tr}(r) = \Phi(r) - \Phi(r_c), \quad r \leq r_c$$
$$= 0 \quad . \quad r > r_c$$

All quantities are calculated in reduced (dimensionless) unit taking '$\varepsilon$', '$\sigma$' and mass of a particle '$m$' as unit of energy, length and mass, respectively.

$$T^* = \frac{k_B T}{\varepsilon}, \quad \rho^* = \rho\sigma^2, \quad \xi^* = \exp(\mu^*/T^*)$$

where $T^*$, $\rho^*$, $\xi^*$ are reduced temperature, reduced density and reduced activity, respectively.

All our studies have been carried out at temperature $T^* = 0.427$. To obtain coexistence activity at this temperature for each different cut-off radius we have performed GC-TMMC[11] simulation studies in a square box of length $L = 25.0\sigma$ near a trial coexistence $\xi^*$. Resulting probability density as a function of density of the system was then extrapolated with respect to $\xi^*$ to obtain the exact coexistence activity[11,12],

$$\ln \Pi(\rho; \mu') = \ln \Pi(\rho; \mu) + \beta \rho V(\mu' - \mu).$$

For different cut-off at their coexistence activity line tensions were obtained carrying out GC-TMMC simulation in a rectangular box such that a rectangular slab is preferably formed at the top of the free energy barrier as a function of density of the system [Fig.1]. Line tension in the thermodynamic limit was calculated following the method suggested by Binder[13,14]. The computed value of line tension for $r_c = 2.5\sigma$ at $T^* = 0.427$ is consistent with the previous value obtained from MD simulation study using a different method[15].

At saturation-ratio, $S = 1.1$ ($S = \xi^*/\xi_0^*$, $\xi_0^*$ is the activity at coexistence) free energy of nucleation as a function of liquid embryo size is obtained from simple Boltzmann Monte Carlo simulation in grand canonical ensemble [Fig.4]. As free energy barriers at this saturation-ratio are not too high, non-Boltzmann sampling method was not employed.

### A. Choice of order parameter

As in CNT the size of a liquid cluster (n) is chosen as the order parameter. To know the size of a liquid cluster first we need to define a 'liquid particle'. To do that we have followed the method developed by Stillinger[16] and later modified by ten Wolde and Frenkel[17]. The liquid particle and liquid cluster are defined as follows:

If any two particles are within a certain distance ($q_c$) then they are considered to be neighbors of each other. The distribution functions of the number of neighbors per particle in the bulk liquid and in the bulk vapor phases at coexistence for $r_c = 7.0\sigma$ are shown in Fig.2. The distributions are well separated near neighbor number 3. From these distributions we have defined a particle as liquid-like if it has more than three neighbors.

After identifying liquid-like particle according to the above definition, we have applied the criterion that any two liquid-like particles which are within the distance $q_c$ belong to the same liquid cluster. So, the size of a liquid cluster is the number of liquid-like particles in that cluster. From radial distribution function of bulk liquid we have seen that the position of the first minimum (i.e., the first shell) is 1.65σ. So, $q_c$ is chosen to be 1.65σ. We have checked that the criteria of liquid-like particle and cluster do not change much with cut-off radius. So, for all cut-off radii the definitions remain the same.

## IV. Results and Discussions

### A. Estimation of the line tension

To compute interfacial vapor-liquid line tension at coexistence the coexistence chemical potential has to be known. Using GC-TMMC method first we have calculated coexistence chemical potential for different cut off radius of potential and then at those coexistence points we have obtained free energy of formation of a vapor-liquid interface[18]. We have applied Binder's method to calculate line tension in thermodynamic limit [Fig.3]. According to this theory the system size dependent line tension can be written as

$$\gamma_L = c_1 \frac{1}{L} + c_2 \frac{\ln L}{L} + \gamma_\infty$$

The interface energy per unit length for different system sizes has been calculated and then extrapolated to obtain $\gamma_\infty$.

According to the best of our knowledge there exists only one simulation study to calculate line tension in 2D system ($r_c$ = 2.5σ, $T^*$ = 0.427, $\gamma^*$ = 0.05)[15]. Our calculated line tension value for $r_c$ = 2.5σ is in good agreement with the previous value. The line tension values for different cut-off are given in Table I.

| $r_c$ | $\gamma^*$ |
|---|---|
| 2.5σ | 0.05(1) |
| 4.0σ | 0.14(9) |
| 7.0σ | 0.17(3) |
| Full LJ | 0.17(8) |

Table I. The vapor-liquid line tension in 2D LJ system for different cut-off values

From the line tension values for different potential cut-off radii, we can see that the growth of line tension with the cut-off values slows down and eventually converges as we go to the larger cut-off. From simulation we have obtained line tension for full

Lennard-Jones potential ($\gamma^* = 0.178$). Earlier DFT calculation[8] taking temperature dependent diameter of particle shows that the surface tension for full LJ system at $T^* = 0.427$ is about 0.32. From our simulation study we have got much lower value.

**B. Free energy surfaces**

We computed the nucleation barriers for $r_c = 2.5\sigma, 4.0\sigma, 7.0\sigma$ and also full LJ potential at saturation-ratio S =1.1. The computed nucleation barriers are shown in Fig.4. From the figure we can see that both the nucleation barrier and the size of the critical cluster increase with an increase in the potential cut-off radius. This is expected as the surface tension increases markedly as we increase the cut-off radius. It has been shown that in 2D LJ system the phase diagram is strongly dependent on the potential cut-off[19]. Here we see that the nucleation phenomenon is also strongly dependent on the cut-off. However, beyond the cut-off radius $7.0\sigma$, there is only a negligible effect of cut-off on nucleation, and we have clearly obtained the converged result.

**C. Comparison with CNT**

By comparing the free energy barriers obtained from simulation with those predicted by CNT (with the surface tension obtained from simulation) at saturation-ratio S = 1.1, we find that the CNT underestimates the nucleation barrier for all cut off at the temperature we have studied. When both the free energy barrier and line tension are calculated with $r_c = 4.0\sigma$ and $7.0\sigma$, the *critical cluster size* prediction of CNT is moderately accurate. However, it fails badly in the case of $r_c = 2.5\sigma$. The critical cluster size predicted by CNT for $r_c = 2.5\sigma$ is much smaller than that obtained from simulation.

In fact, the free energy barrier prediction of CNT is also not good for any of the cut-off we studied at S = 1.1. We compute the rate of nucleation both as predicted by CNT and as observed from simulation using the following relation:

$$J = J_0 e^{-\Delta\Omega^*/k_B T}$$

Where, $J_0$ is the prefactor used in CNT. We assume the same value for our simulations and the ratio $J_{CNT}/J_{sim}$ is plotted in Fig.8 for different cut-off values and it can be observed that $J_{CNT}$ over-estimates $J_{sim}$ by as much as ~650, for all of the cut-off values. This is in disagreement with the theoretical estimate of Zeng who found from DFT calculation that $J_{CNT}$ and $J_{sim}$ differs at most by a factor of 2, at $T^* = 0.427$ for saturation-ratio corresponding to $J_{CNT} = 1\ cm^{-2}s^{-1}$. One of the reasons for the over-estimation of rate (and under-estimation of barrier) by the theory (compared to simulation result) is that Zeng used the theoretical value (obtained from DFT) of line tension equal to 0.32 (in reduced unit) where as the converged value is 0.178. This itself increases the height of the barrier by ~80% over what would be obtained by using the right value of the line tension!

Interestingly, whereas in 2D system, the CNT under-estimates the free energy barrier, it over-estimates the barrier in 3D system[17]! However, in the case of 3D nucleation for a range of saturation-ratio starting from S = 1.53 to 2.24, there is a relatively small, nearly constant offset of nucleation free energy barrier ($\sim 5k_B T$) between CNT prediction and computer simulation result. In 2D even close to the coexistence, the CNT underestimates the nucleation barrier by $\sim 6.5 k_B T$ which is about 70% smaller than the correct value.

We have studied the shapes of the critical clusters by taking snapshots of critical clusters. It shows that the clusters are deviated from circular shape [Fig.7]. For large clusters it has been found that there are voids present inside the liquid clusters. We attempted to do shape correction on CNT at $r_c$ =2.5σ for S = 1.08 as the cluster is not circular [Fig.5]. It has been found that in case of crystallization in 3D LJ system the modification of the shape of the critical cluster, assuming it as ellipsoid, improves the nucleation free energy barrier considerably[20]. It also improves the size of the critical cluster. The cluster is assumed to be elliptical in shape and then from the positions of all the particles in the cluster the aspect ratio is calculated from which the periphery of that ellipse has been obtained. After this modification the nucleation free energy increases a little but the critical cluster size gets improved a lot. We also calculated the length of the periphery using another method. A circle is rolled over the cluster and from the distance covered by circle the circumference is calculated. In this method the improvement of the free energy barrier is better than ellipse assumption but still it markedly differs from simulated free energy barrier. Although the improvement of the nucleation barrier is not satisfactory, the size of the critical cluster is good.

## V. Conclusion

Because of the presence of large scale density fluctuations, phase transitions in two dimensions is known to show features which are often dramatically different from those in three dimensions. Because of the smaller number of neighbors in 2D, it is difficult to stabilize an interface between a gas and a liquid, leading to large fluctuations of the interface. According to the capillary wave approximation, the width (w) of the interface grows as $\sqrt{L}$, where L is the length of the interface. In 3D, this growth is logarithmic in L. These fluctuations manifest here in terms of a strong dependence of the properties of critical nucleus on the cut-off of the Lennard-Jones potential. One needs to use a cut-off as large as 7σ to obtain reasonable results. It has already been observed by Smit and Frenkel that one may need to use the full range of the potential in 2D to obtain reliable phase diagram. We also find that the surface tension is sensitive to the cut-off.

There are several results obtained here which are of considerable interest. First, we find that the free energy barrier of nucleation is much larger than the prediction of the classical nucleation theory, as shown by Fig.6. Second, the size of the cluster is also vastly underestimated by CNT. In both the two results, we have used the accurate value of surface tension obtained by computer simulation. The third important result is the value of the surface tension which is about 44% less than the available theoretical estimate of Zeng. It should be pointed out that Zeng was absolutely correct in observing that the then available estimate of surface tension ($\gamma^* = 0.05$, calculated with a cut-off of only 2.5σ) should be much smaller than the actual value ($\gamma^* = 0.178$). But the value used by Zeng was too large ($\gamma^* = 0.32$ instead of 0.178). We conjecture that this overestimate by the DFT theory of the surface tension might have led Zeng to conclude that CNT could be more accurate for 2D than for 3D. At saturation-ratio S=1.1, we find that the cluster shape is nearly spherical with large or full cut-off.

The reason for the under-estimation of the barrier height by the CNT for the 2D gas-liquid nucleation can perhaps be understood in terms of the prevalence of the non-spherical shape of the critical cluster (see Fig.7). Although this non-spherical nature of the critical cluster decreases with the increase in the LJ interaction cut-off radius, yet a substantial degree of non-sphericity remains even when the full range of interaction is used in simulations. Because of the non-sphericity of the cluster size, the contribution of the line tension term increases which gives rise to a barrier height larger than the CNT which assumes a circular shape. We have shown in Fig. 5 that inclusion of the enhanced line due to non-circular nature of the critical cluster increases the barrier height. Another possible reason is that the liquid phase that nucleates initially is of lower stability (higher free energy) than the bulk liquid (at the same thermodynamic conditions) as it contains voids. This is similar to the case of liquid-crystal nucleation where the solid that first nucleates is more disordered than the bulk solid. Therefore, the free energy difference $\Delta\Omega$ used by CNT in Eqs. 1, 3 is less for the critical cluster. This effect could be particularly acute in 2D where fluctuation effects are strong, leading to less order in the critical liquid cluster.

Our preliminary study shows that the spinodal of gas-liquid transition in 2D is at a smaller saturation-ratio than in 3D (where $S_{spinodal}$ ~2.4-2.5 with LJ interaction. It will be interesting to study the nucleation behavior in 2D near the spinodal, although we expect the fluctuations will make such a study quite challenging.

**Acknowledgement**

This work has been supported in part by grants from DST (India). S.C. acknowledges CSIR, India for a research fellowship. B. Bagchi thanks DST for support through J.C. Bose Fellowship.

# Figures

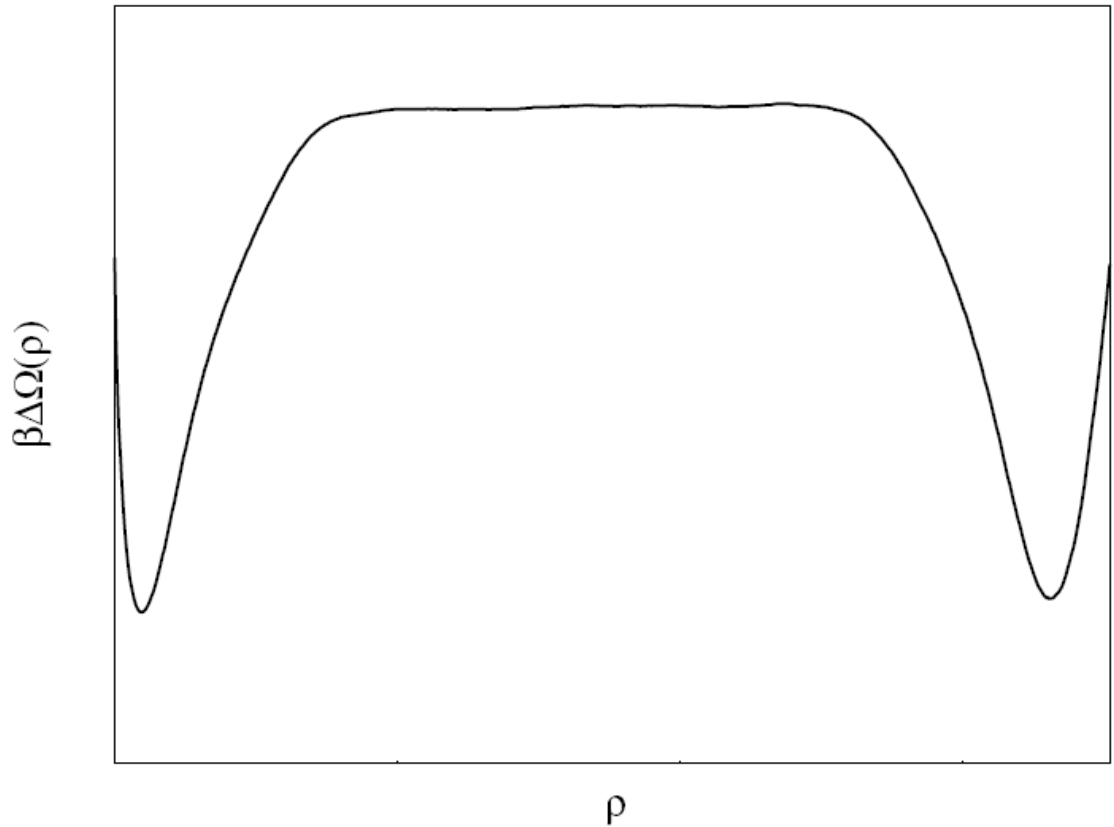

Fig.1: The calculated free energy of formation of a liquid slab with two parallel liquid-vapor interfaces. 'ρ' is the density of the entire fluid. See Ref.14 for details.

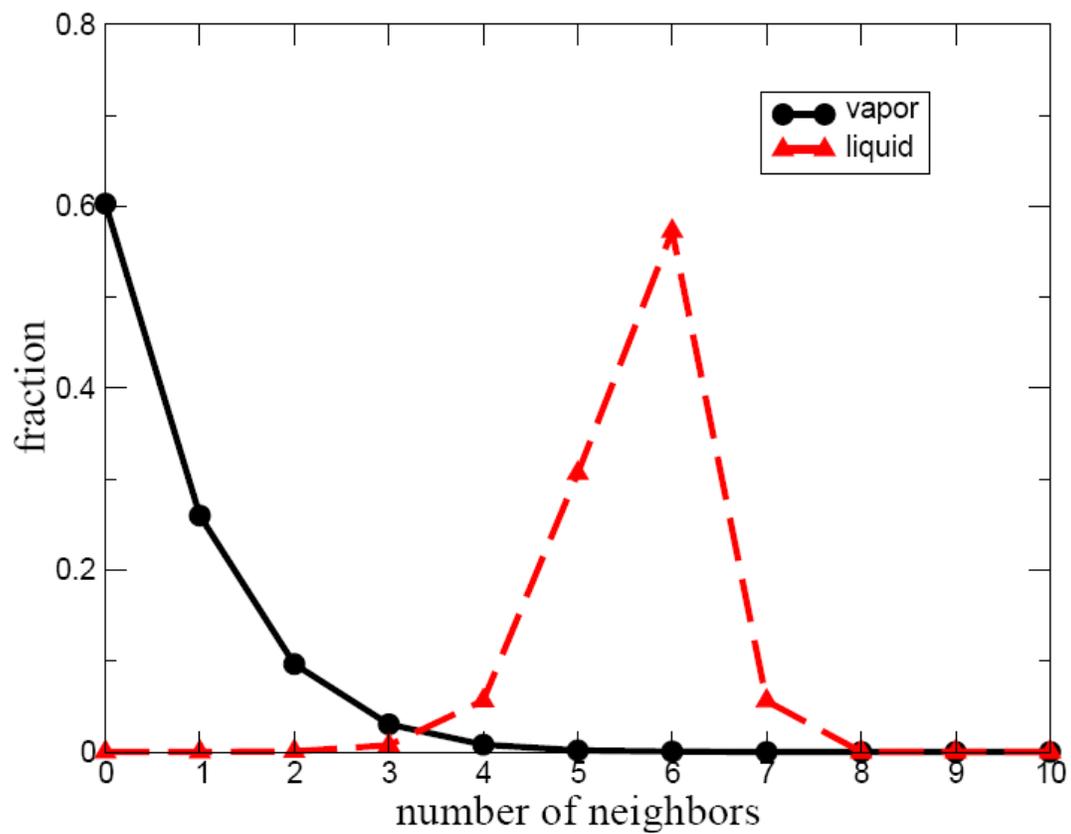

Fig.2: The distribution of the number of neighbors of a particle in bulk vapor and in bulk liquid phases at coexistence. Here $T^*=0.427$, $r_c=7.0\sigma$.

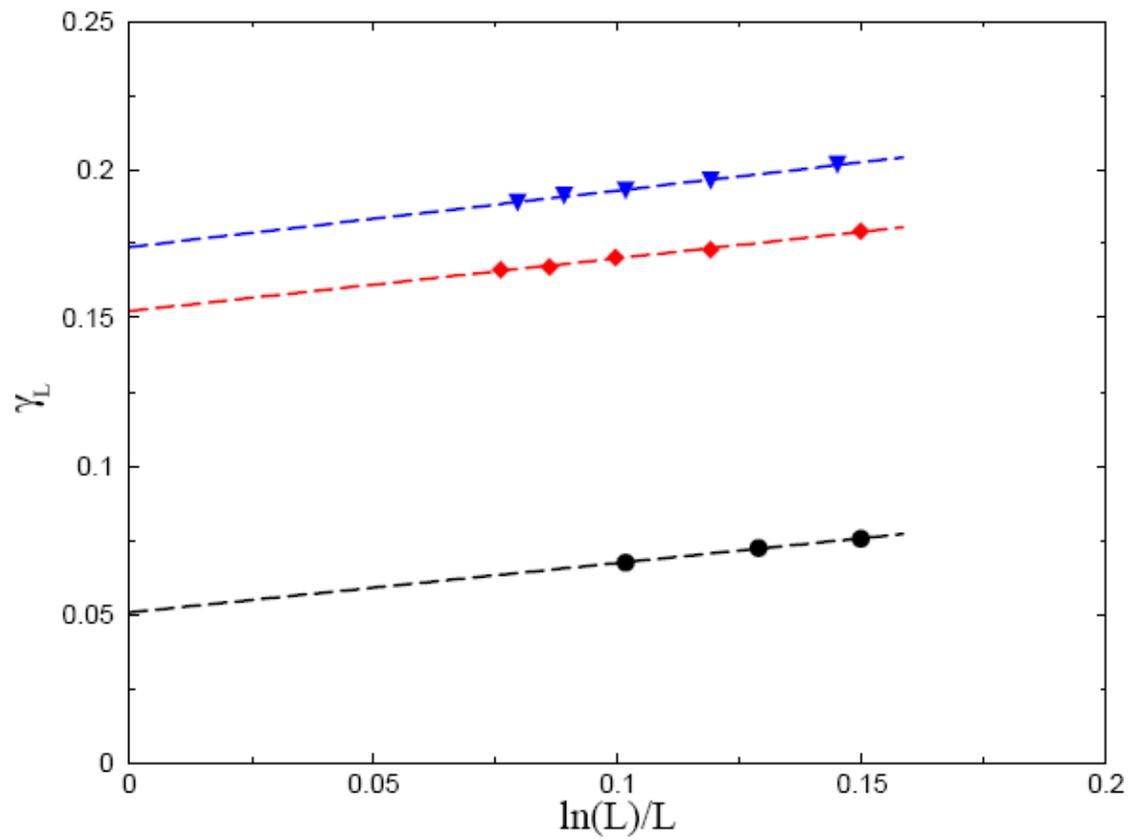

Fig.3: The liquid-vapor interface energy per unit interface length vs. ln(L)/L where L is the length of the interface. In the figure circle, diamond and triangle up are for $r_c = 2.5\sigma$, $4.0\sigma$ and $7.0\sigma$, respectively. Dotted lines are corresponding linear fits.

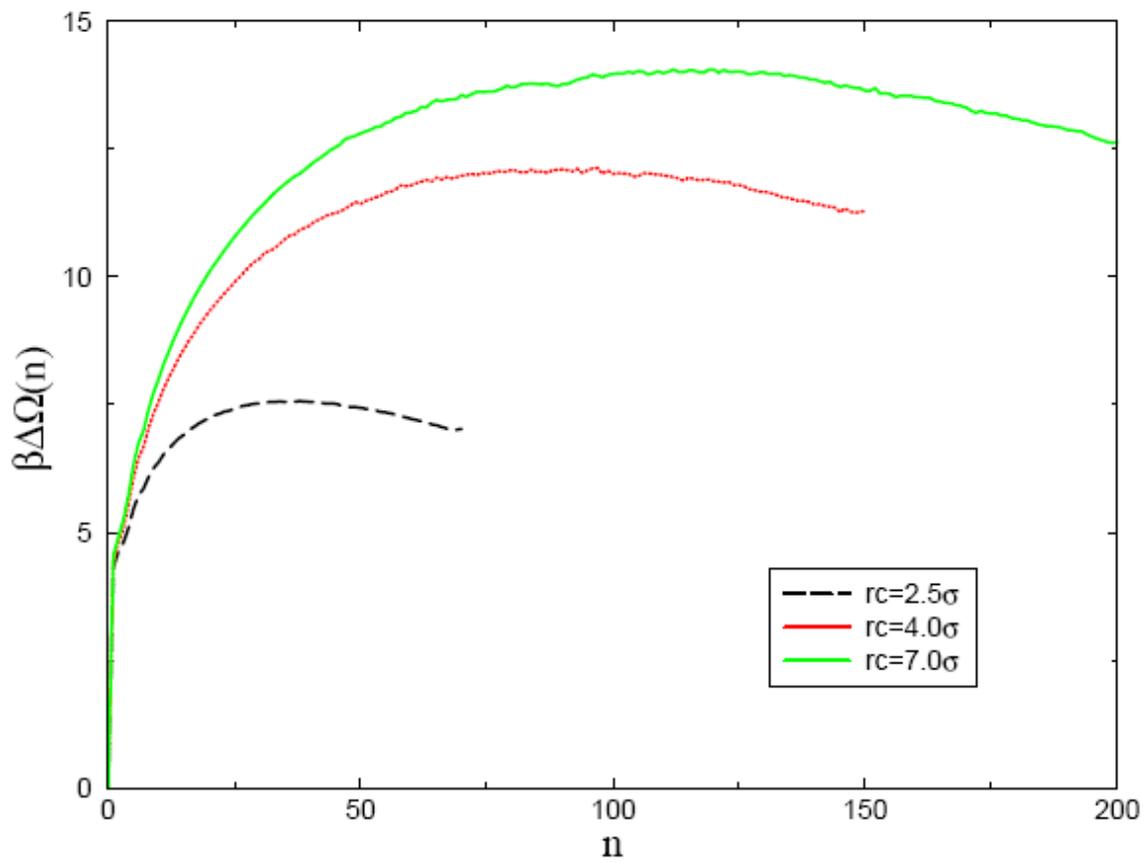

Fig.4: The computed free energy of formation of a liquid embryo as a function of size of the embryo. Here n is the number of atoms in the cluster. Saturation-ratio S =1.1, $T^*$ = 0.427.

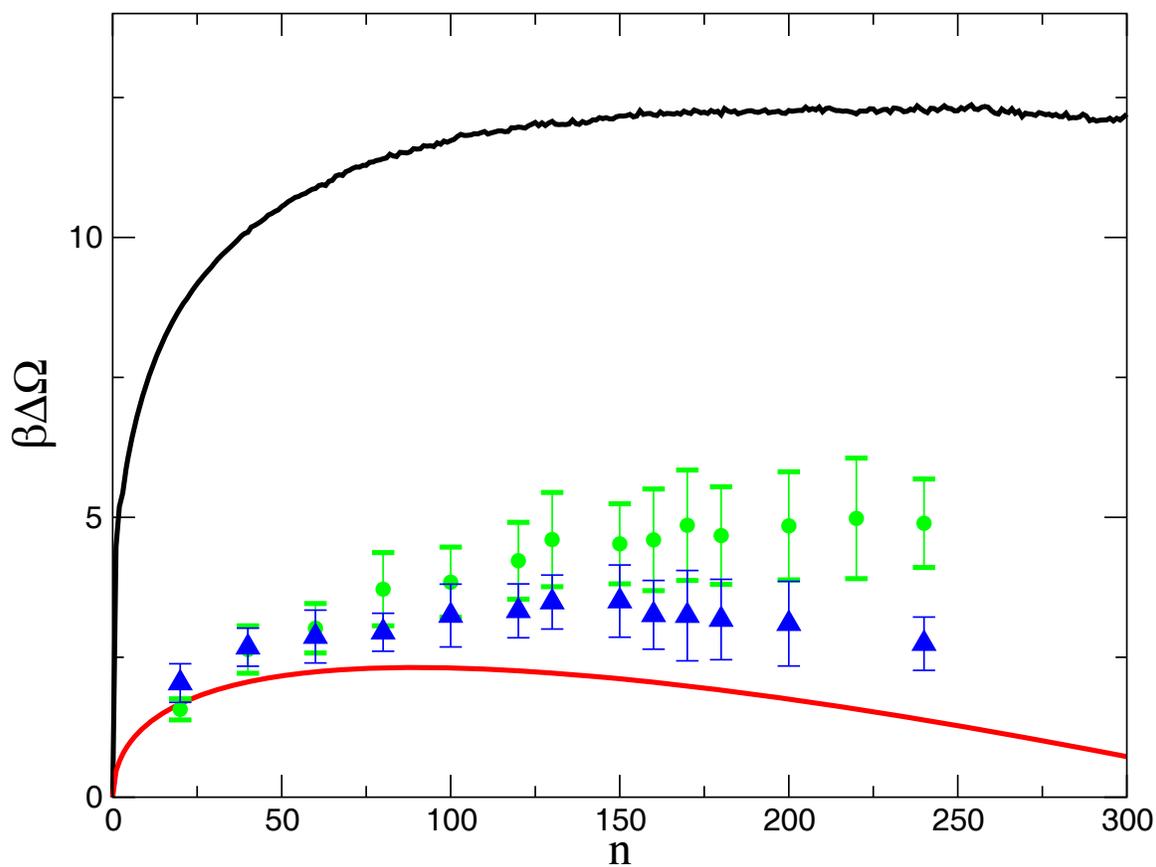

Fig.5: The free energy of nucleation versus size of a liquid cluster. Black line is from simulation, red line is CNT calculation, triangle up is after shape correction approximating shape of the cluster as an ellipse. Circle represents shape correction from rolling method. Here $T^*$=0.427, $r_c$=2.5, S=1.08.

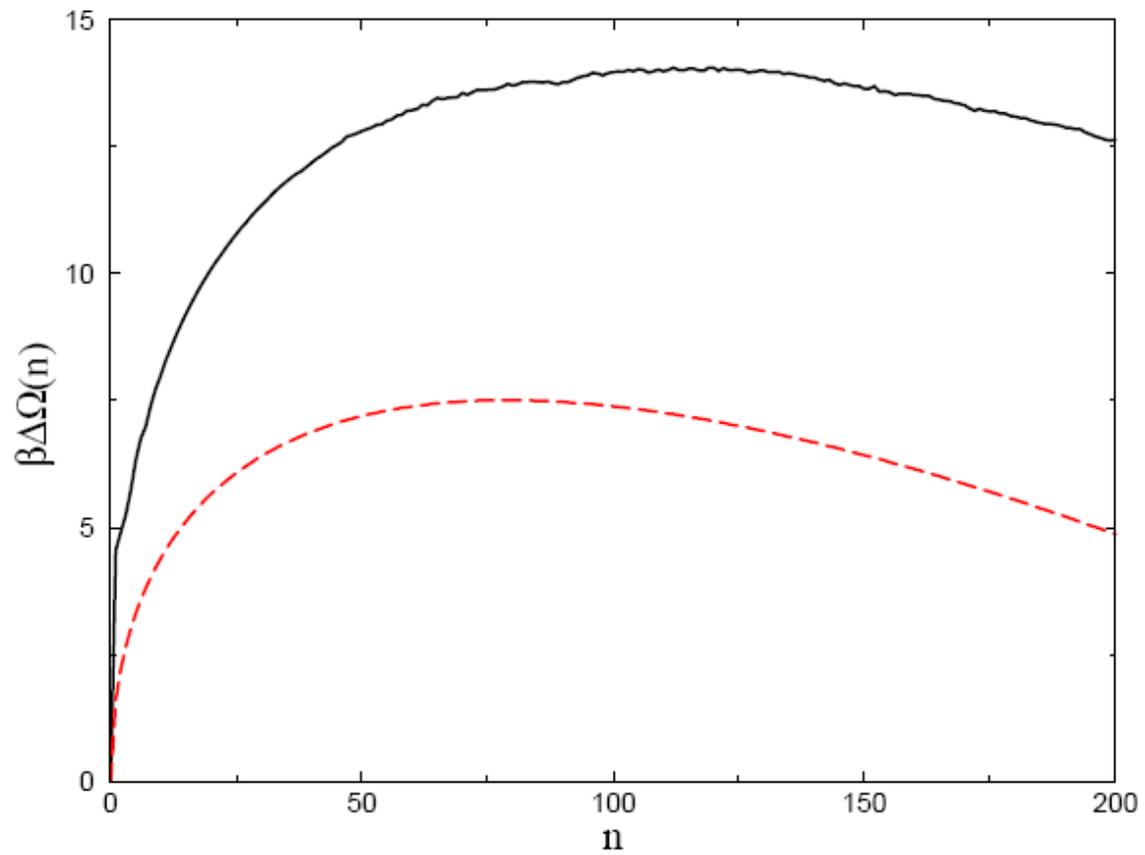

Fig.6: The comparison of the computed nucleation free energy barrier with CNT prediction for $r_c$ = 7.0$\sigma$, S = 1.1, $T^*$ = 0.427. Solid line is the simulation result and the dotted line is the CNT prediction. The free energy barrier calculated with full interaction range is almost indistinguishable from $r_c$=7.0$\sigma$.

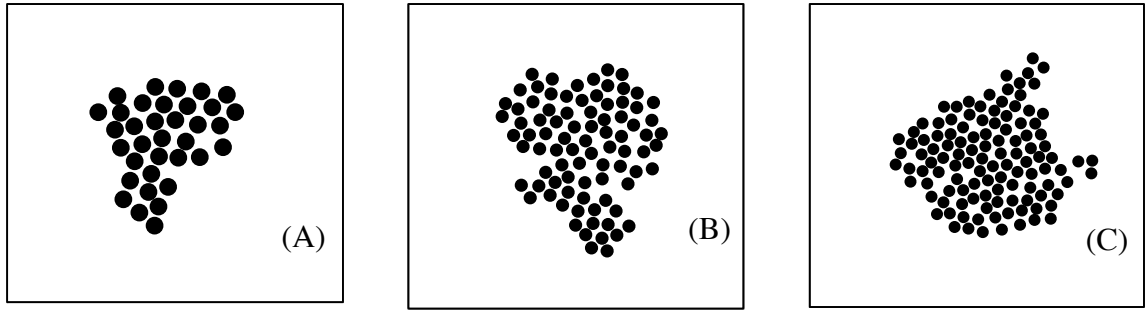

Fig.7: Snapshots of critical clusters at S = 1.1, T* = 0.427 with the following cut-off radii. (A) $r_c$ = 2.5σ, size of the critical cluster $n^*$ = 35. (B) $r_c$ = 4.0σ, $n^*$ = 85. (C) $r_c$ = 7.0σ, $n^*$ = 115.

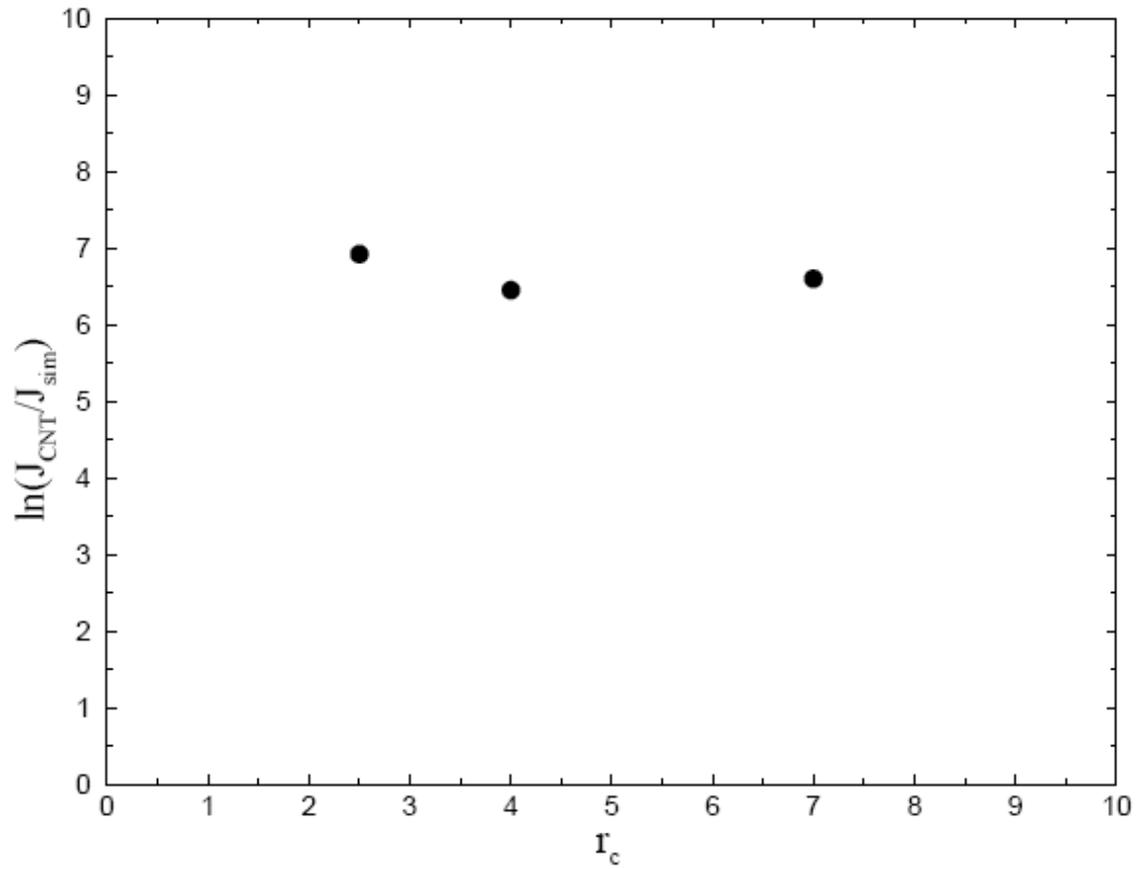

Fig.8: The comparison of rate of nucleation for different cut off taking pre-factor as the same as in CNT. Here $T^* = 0.427$, $S = 1.1$. Note that although the free energy barrier and the critical cluster size change with the cut-off, the ratio $J_{CNT}/J_{sim}$ is nearly independent of the cut-off.